# Privacy, Trust and Identity in Pervasive Computing: A Review of Technical Challenges and Future Research Directions

Ameera Al-Karkhi[1], Adil Al-Yasiri[2] and Nigel Linge[3]

[1,2,3] Department of Computing, Science and Engineering, University of Salford, Manchester, UK

```
a.a.s.al-karkhi@edu.salford.ac.uk
   a.al-yasiri@salford.ac.uk
      n.linge@salford.ac.uk
```

**ABSTRACT**

*Developments in pervasive computing introduced a new world of computing where networked processors embedded and distributed in everyday objects communicating with each other over wireless links. Computers in such environments work in the background while establishing connections among them dynamically and hence will be less visible and intrusive. Such a vision raises questions about how to manage issues like privacy, trust and identity in those environments. In this paper, we review the technical challenges that face pervasive computing environments in relation to each of these issues. We then present a number of security related considerations and use them as a basis for comparison between pervasive and traditional computing. We will argue that these considerations pose particular concerns and challenges to the design and implementation of pervasive environments which are different to those usually found in traditional computing environments. To address these concerns and challenges, further research is needed. We will present a number of directions and topics for possible future research with respect to each of the three issues.*

## 1. INTRODUCTION.

The wide development and integration of sensing, communication and computing have led to the development of pervasive computing, which offers the distribution of computational services within environments where people live, work or socialise. There are advantages in implementing such environments such as, moving interaction with computers out of a person's central focus and into the user's peripheral attention where they can be used subconsciously. Another advantage of pervasive computing environments is to make life more comfortable by providing device mobility and a digital infrastructure that has the ability to provide useful services to people in the environment, when and where they need them. It is common that a user in these environments will maintain various connections with many smart devices regardless of the hardware specifications or the software restrictions. Such devices collectively participate in the provision of the required service without the conscious or explicit knowledge of the user as stated by Weiser [1]. However, at the same time pervasive computing presents many risks and security related issues that were not previously encountered in more traditional computing environments. In particular, issues such as privacy, trust and identity become more challenging to the designers of such environments. Designing secure pervasive environments requires the system to reliably and confidently identify the user who wishes to access the environment's resources. It is also important to appreciate the risks involved in establishing and verifying the identity of users in such environments. Privacy is also important as users need to be confident that their personal information is not used in a way that they do not approve of. Privacy in such environments is particularly important as the system needs to be protective of the users' data and perceived by the user to be that way. Trust within such systems presents another challenge due to the fact that trust relationships are much more complex than those normally found in more traditional environments. In pervasive environments it is very difficult to define the boundary of trust domains, which is important when defining trust relationships. Trust is also important when users often cross such boundaries and therefore normal authentication procedures may not be practical. This paper reviews the technical advances and challenges with respect to each of these issues within pervasive computing. In section 2, the paper discusses the network access issues in pervasive computing and how they differ from those in traditional computing. It presents the views of various authors about how accessing the network resources may be handled in pervasive computing. The following three sections review previous research related to privacy, trust and identity within pervasive computing. In section 7, we provide a comparison that summarises the main differences between pervasive computing and more traditional systems in relation to the challenges faced in designing secure pervasive computing environments. In section 8, the paper provides a number of suggested topics for future research in each of privacy, trust and identity. The paper is then concluded by a summary section.

## 2. Network Access in Pervasive Computing

In pervasive computing environments, users expect to access resources and services anytime and anywhere, leading to different and more serious types of risks. A network inside such environments is expected to dynamically connect to other networks and change its topology. For example Zugenmaier and Walter [2] stated that when mobile devices join and leave a certain mobile zone, and as their wireless short range radio interface goes into and out of range of access points or other mobile devices, using a traditional technique such as a firewall will be inadequate because a separate firewall is needed to protect each of the devices. They [2] explained that in today's world of networking and computing, there is no problem in protecting systems because network resources can be policed using firewalls or intrusion detection systems to separate trusted and non-trusted parties. However, in pervasive environments these techniques are unusable and unworkable, especially when considering the case of a device stolen or lost. According to Weiser [1], pervasive computing objects can be divided into two main groups:

- Personal devices which are usually carried by individuals,
- Infrastructure devices which are embedded in the environment.

The interaction between these two categories will define the needs for a new resource access model. Therefore Zugenmaier and Walter [2] proposed an ideal state in which the new access model will not be achieved by forbidding everything; but by "monitoring, evidence gathering and reconciliation". They discussed how to build a new framework which includes a technical security solution, services and rules for good behaviour and ways of dealing with pervasive computing security breaches.

User authentication and access control strategies can be used to provide security in small networks and stand alone computers. These strategies are used to verify the identity of a person or a process in order to enable/restrict the ability to change, use, or view resources. However, the wide development and flexibility in distributed networks, such as the Internet and pervasive computing environments, show that these strategies are inadequate because such systems lack central control and their users are not all predetermined, leading to serious risks and access control problems. Consequently, security is a crucial design issue in pervasive computing (because of the usability and expansion of pervasive computing applications) which depends on the security and reliability provided by the applications [3].

In a pervasive environment, there is a strong possibility that people will be monitored by a large number of invisible computers in every field of their life, either private or public. The work in [4] suggested that the designer of those systems should understand how people can trust such an environment and then accept it. Furthermore, pervasive computing systems should consider other issues such as privacy, trust, and identity. Because of the wide interaction between the pervasive environments and people, privacy becomes particularly important as people learn about the existence of such systems and become protective of their own privacy. This is because they do not know how their personal data is being collected and what the purpose behind this is. Therefore, many researchers suggest different ways to protect the users' privacy, such as anonymity, and giving the right to users to choose whether to distribute and exchange their personal data or not. Whenever a user trusts a system, he/she will be more inclined to reveal their personal data. Because of the dynamic connections between the device and user, the user will depend on the trust relationship agreed between them. Establishing user identity in pervasive computing environments requires special attention because using the traditional techniques will be insufficient to establish and verify the identity of users. This is due to the mobility of devices and the random connection between them and users. Thus, providing a method to verify the real identity of a user will be necessary for the success of such an environment. Moreover, it is important to note that pervasive computing environments require a new type of authentication (authentication of artefacts), which means a physical artefact has to prove that it knows a secret, in addition to the need for other traditional types of authentication as stated by Bussard and Roudier [5].

## 3. Privacy in Pervasive Computing

In pervasive computing environments, where the concentrations of 'invisible' computing devices are continuously gathering personal data and deriving user context, the user should rightly be concerned with their privacy. Devices may reveal and exchange personal information (such as identity, preferences, role, etc) between smart artefacts in pervasive systems. In a context where devices cannot be assumed to belong to a single trusted domain, privacy becomes a major issue. It is crucial to develop and create privacy-sensitive services in pervasive computing systems to maximize the real benefit of such technologies and reduce feasible and actual risks. Because such systems collect a huge amount of personal information (such as e-mail address, location, shopping history... etc) and because people are typically concerned about their personal information, it is conceivable that they will be reluctant to participate in pervasive environments. Thus, it is paramount to provide a mechanism that ensures privacy is maintained at all times. Privacy can be defined, according to Steffen et al. [6], as "an entity's ability to control the availability and exposure of information about itself". In [7], the authors identify five characteristics that make such systems very different from today's data collection systems, which are:

1. new coverage of smart environments and objects will be presented everywhere in our life;
2. data collection will be invisible and unnoticeable;
3. the collected data will be more intimate than ever before; for example how people feel while doing something;
4. the underlying motivation behind the data collection;

5. the increasing interconnectivity which is necessary for smart devices to cooperate in order to provide a service to users; this results in a new level of data sharing making unwanted information flows much more possible.

Together, these characteristics indicate that data collection in the age of pervasive computing is not only a quantitative change from today, but also a qualitative change. Users in pervasive computing environments do not know what is done with their personal information and a service may store or process the provided data in some way that is not intended by the user. This fear makes people feel more concerned about their privacy.

Research presented in [8] showed that over the past years, a range of various websites like social networking services and photo and video sharing services, put a demand for users to share their information with each other. When using such services, users demand control over the conditions under which their information is shared. The research found that more complex privacy setting types can lead to more sharing. For example, Facebook started to head to more complex privacy-setting types. This suggested that offering comfortable and flexible privacy settings could make services more valuable. The work presented a three-week study in which locations of 27 participants have been tracked and participants were asked to rate when, where and with whom they would be comfortable to share their locations. The research studied location-privacy preferences according to the case study. When a participant visited a location, they were asked whether or not he/she is willing to share this information with each of four different groups: close friends and family, Facebook friends, the university community and advertisers.

As Weiser [1] noted, "If the computational system is invisible as well as extensive, it becomes hard to know what is controlling what, what is connected to what, where information is flowing, how it is being used and what are the consequences of any given action". Then he [1] referred to privacy as a solution research issue; it has always been raised as a crucial issue for the long-term success of pervasive computing. The concept of privacy has become one of the main concerns as the technology of smart artefacts develops. Moreover, in the developed world there has also been a growing awareness of privacy issues in general, particularly due to the increased use of the World Wide Web. Weiser [1] stated that a well-designed pervasive system should eliminate the need for giving out some items of personal information. For example, schemes based on "digital pseudonyms" could eliminate the need to give out items of personal information that are routinely entrusted to the network's today, such as a credit card number and an address. Langheinrich [9] stated "Everything we say, do, or even feel, could be digitized, stored, and retrieved anytime later. We may not (yet) be able to tap into our thoughts, but all other recording capabilities might make more than up for that lack of data." The author formulated six principles for directing system design based on a set of fair information practices common in most privacy legislation. The principles are: Notice, Choice and Consent, Proximity and Locality, Anonymity and Pseudonymity, Security and Access and Recourse. In another publication, Langheinrich [10] considered designing a perfect mechanism for protecting privacy would be difficult to achieve. Therefore he proposed a system where the users are allowed to be alerted about their privacy. The system relies on social and legal principles from real life, instead of designing a system to ask other people to respect the user's privacy. This system, named privacy awareness system (*pawS*), permits data collectors to process personal data and management policies, and to describe tools for manipulation of personal information (storing, deleting and modifying information). In the main, this system is based on four of the above six principles: (notice, choice and consent, proximity and locality, and access and recourse), while the other two principles (Anonymity and Pseudonymity, and Security) are useful tools and a supportive part of the infrastructure. The developed pawS architecture (Privacy Preferences Project P3P) includes two main parts: privacy proxies and a privacy-aware database.

1. Privacy proxies: In P3P, privacy proxies permit the automated exchange and update of both privacy policies and user data. It is implemented as a group of services using Simple Object Access Protocol (SOAP), running on a web server. As an example, a user wishes to access a specific service (print service or location tracking by video camera) where some personal information is required by the service to provide its function. The service will then contact the service proxy who provides a list of the available platforms, the different levels of services and the required data for each level. The P3P allows users to set their own personal privacy preferences and reply with the required information. When the communication is successful, the service proxy replies with an agreement ID which is reserved by the user proxy for reference. This ID can be used by the user at any time to authenticate him/her and find, update or delete any personal information, using HTTP over SSL to prevent eavesdropping.

2. Privacy-aware database (called pawDB): This combines the users' collected data elements and their privacy policies into a single component of storage which then handles the data according to its usage policy. It uses the following algorithm:

    i. first the (P3P) policies, which describe why the data is collected, are imported into relational tables using XML-DBMS and are assigned a reference number;
    ii. using the API, the data will be inserted into the pawDB;
    iii. the system compares the submitted data with its privacy policy governing it and transparently stores all the data and their privacy policy together;
    iv. to query any of the stored data, the user should describe what they are, the purpose of their query and for how long they will keep their information;
    v. pawDB compares each query and its usage policy with the data collection policy of each individual element and transparently withholds a particular piece of information in case of a mismatch between the two.

According to the above, pawS is a personal privacy assistant which keeps track of the collected data with and without the user's help. It assists users to enable or disable a service depending on their preference. To protect users' privacy, many researchers suggest Anonymity which means personal identity or personally identifiable information (of a person) is unknown; in other words, the user uses the system without any identification or identifier that distinguishes them from other users. Zugenmaier et al. [11] introduced a new attacker model, called the Freiburg Privacy Diamond model (FPD), to analyse anonymity mechanisms with regard to device mobility.

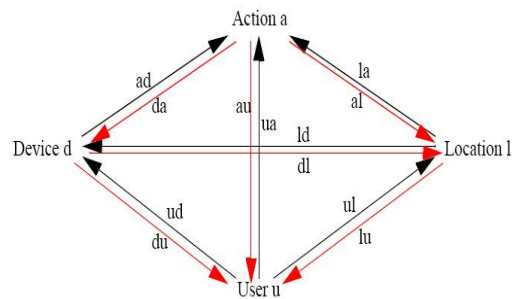

Figure1. The privacy diamond model [12]

This FPD model enables an analyser to adjust the "strength of the attacker in a fine grained way detailed and dependent on the computing environment". It can be used to understand the issues affecting anonymity in a communications environment. Their model uses four types of entities to characterise information about anonymity, which are: the performed action, the *device* used for performing the action, the user who performs the action and the location of the device and the user, as shown in Figure 1. The authors described how these entities are related to each other and how an attacker can use familiarity with these entities including relationships among them to break anonymity.

Zugenmaier and Hohl [12] used the FPD Model to analyse various mechanisms for ensuring anonymity in pervasive computing and highlighted the problems that arose out of a model called "the one user many devices model". So, an extended privacy diamond was introduced, which may be used to examine if a user is anonymous when using more than one device per user. The results of the single FPD model are transferred to the extended privacy diamond.

Hajime et al. [13] showed that because any entity (object/person) in IT systems has a fixed identifier (ID) such as radio frequency identifier (RFID), which can be used for identifying someone, there is a possibility that revelation (disclosure) of a fixed ID will cause privacy problems. So, they proposed a method called "identity control" to conceal a fixed ID during transmission times and set the level of switching to reveal the real name and pseudonym from the anonym according to the authority of the receiver and the wishes of the sender. Moreover, according to this method the real name of the sender is changed into a pseudonym, and the pseudonym is changed into an anonym using encryption. Then the receiver can either decrypt the received anonym to real name or pseudonym, or use it as is. The authors claim that the sender can control the way real names and pseudonyms are deciphered (by the receiver) by changing the encryption key. This way, the sender can conceal job-related information from a particular receiver by making limited key information available to them.

Chatfield and Hexel [14] studied user identity and privacy in smart environments, and suggested a possible solution to the management of user identity and maintenance of user privacy within such environments. They also described the concept of "User Selected Pseudonyms", as a possible solution to identity management in intelligent environments. It is a method for allowing users to manage their identity in pervasive computing environments, by granting them control (either directly or through privacy preferences) to decide what information about themselves they share with an environment. They used this method to allow users to control the risks to their privacy and give them confidence to utilise the services provided by these environments while maintaining the same anonymity currently enjoyed without them. The main idea behind this method was that, when users enter a smart environment, they have the option of providing the environment with a previously used pseudonym (or history of interaction) using their handheld device, to create a new pseudonym or to remain anonymous. Users can keep many different environmental pseudonyms, allowing them to effectively choose their required level of anonymity at any given time. This choice allows users flexibility in using these environments as they wish, allowing them to restrict sharing information about themselves to situations when a reason is given to do so. This transparency allows users more control over their privacy, and potentially greater confidence when interacting with these environments. In addition, Alfred and Jörg [15] identified the privacy benefits of using pseudonyms within adaptive systems, and suggested that users can be allowed to adopt multiple pseudonyms to improve their privacy. This research examined users' interactions with intelligent environments, and sought to identify an interaction technique to improve users' experience within such environments in relation to privacy and confidence. Allowing users to be anonymous in smart environments and improve their privacy can make system adaptability much less effective. Using pseudonyms has been identified as a method of allowing users to use adaptive systems, while maintaining some anonymity. The research also proposed an architecture for privacy and security which allows users to benefit from personalisation while hiding their identities. They referred to it as a reference model for pseudonymous and secure user modelling. It included a permissions server for role based access control, a way for hiding users' identity, modelling servers, secure transport, a certificate directory, and a reference monitor that controls the access of clients to user models located in the user modelling server.

Privacy protection remains a serious barrier to the widespread deployment of Pervasive Computing environments. Researchers are considering identifying applications and seeking ways for creating interactions that are effective in helping end-users manage their privacy in pervasive computing. Jason and James [16] developed a toolkit (called Confab) for helping the development of privacy-sensitive pervasive computing applications. It provides basic support for building pervasive computing applications, a framework and several customizable privacy mechanisms. In this framework all the personal information of a user will be captured, stored and processed on the user's computer as much as possible, and then the user can control what information to share with others. They focused on authorising people with choice and informed permission, so that they can share the right information with the right people and services in the right situations. A number of researchers have worked on another aspect related to privacy, which concerns monitoring users' behaviour. Within pervasive computing, monitoring capabilities can be intrusive because there are sensors and machines which take over the role of the watchers and begin to store more and more aspects of our daily routine. Because it is difficult to know when people become conscious that they have been monitored and their privacy has been violated, Langheinrich [17] described an approach called privacy boundaries. This approach tries to capture the various reasons a certain flow of personal information is perceived threatening, and then assesses how pervasive computing affects it. The authors also tried to identify and motivate key concepts in personal privacy that should influence the "design and implementation of privacy-aware pervasive computing systems, which are the systems that take the social fabric of everyday life into account and try to prevent unintended personal border crossings". For example, Rhodes [18] presented the wearable memory amplifier, allowing its wearer to continuously record events of their daily life (multimedia diary), which helps them to remember a lot of small details to provide a useful service. There is, however, a cost in increasing the risk at the privacy boundaries.

Another piece of information that is considered sensitive and related to privacy is the user's location information. Hengartner, U., & Steenkiste [19] viewed location information as a sensitive piece of information that should not be disclosed to anyone. They analyzed some requirements for location policies and implemented an access control system that supports flexible location policies, by allowing people to specify their location policy which states who should be allowed to locate them. They implemented a prototype that supports two-fold policy; user policy and room policy. In user policy, the user can state only a building name to be returned and in room policy, the number of people in a location can be returned instead of their identity. Moreover, users can specify their own policies by issuing appropriate digital certificates (used to express their location policy) using a central authority. Also people can choose to share the right information with the right people and service in the right situation.

The work in [20] presented multiple techniques for personal privacy management in pervasive sensor networks. It provided a user-centric privacy protected platform for the deployment of an invasive sensor network. The study also evaluated the usability of an active privacy badge system and the likelihood of using this system as a building-wide privacy protection facility. Throughout the research, results showed that an active badge system for privacy control is the most acceptable method among all the tested choices (disabling data transmission from an active badge system, on/off switches, or the touch screen displays). The results from tests also suggested that if residents of moderately denser buildings block data transmission, the availability of the sensor network will be compromised. Consequently, it is vital to discover a balance between protecting privacy and maintaining enough data flow for the value-added applications employing the network at the same time.

As discussed in this section, many researchers focused their effort towards providing technical solutions to address the privacy concerns which have become more acute in pervasive computing environments compared to traditional computing systems. In such environments, users interact with the surrounding digital devices and have an appreciation of the fact that their personal information is being acquired and revealed to these devices. However, they are unaware where and when this information is collected and how such devices acquire and use it. This situation introduces requirements for users to trust such environments and for clear policies for the exchange of such information to be defined. For example, policies enable the users to keep their information in a secure place without saving any copy of their information in any place. Along with articles covering privacy aspects, we believe that future research will continue this development in providing privacy protection in the physical sensor layer for other forms of personal contextual information. In order to build pervasive computing systems that will be considerate to the privacy of individual users, it is necessary to recognize when people feel their privacy has been invaded.

## 4. Trust in Pervasive Computing

Trust is a vital component in both traditional and pervasive computing environments. It can be defined according to Grandison and Sloman [21] as a "relationship between two entities; trustor (the subject that trusts a target entity) and trustee (the entity that is trusted)". In this section we review research and discuss various methods used to solve the problem of trust in pervasive computing environments. In pervasive environments a large amount of personal information will be collected and as a result of this, many people will not engage in such environments because they do not like to be tracked. According to this, Wagealla et al. [22] used trust to protect users' privacy based on trustworthy information received and by allowing users to decide the amount of information that can be disclosed to them. Their approach relied on dividing users into information owners (users who are tracked) and information receivers (users who like to use the sensed location information). Information owners can therefore specify their own policy in recognition of the fact that people have different attitudes towards their privacy. Access to information is controlled by the information owner, which is expressed in terms of the trustworthiness of the information receivers. The researchers concluded that the privacy of information depends on the level of trust between the information owners and information receivers. The authors introduced a trust-based security model for the collaboration between devices within pervasive computing environments. The model ensures secure interaction between smart devices and services, by addressing the concerns of security and trust. In their work, they identified two security problems in collaborative environments of smart artefacts:

- ensuring the accuracy of personal information, and
- establishing trust between personal and public artefacts to support collaborative tasks

Moreover, the two main sources of trust information (about an entity) are personal observation of the entity behaviour and the recommendation from trusted third parties. The trust information of a particular entity may be stored as historical data. Then, its trustworthiness is assessed before deciding to interact with another entity when risk assessment is undertaken. The results of the risk assessment are change depending on how much is known about the entity's actions in the past. Research presented in [23][24] and [25] suggested a way to increase security by the addition of trust, which is similar to the way security is handled in human societies. They argued that a chain of trust will allow greater flexibility in designing policies and providing more control over accessing services and information in pervasive computing. The proposed solution by Kagal et al. [24] [25] was based on developing a framework called "Centaurus" to create Smart Spaces which includes a message-based transport protocol that is designed to do well in low-bandwidth networks. They based their work on a smart office scenario where a mobile user can access all the existing resources via a handheld device connected over a short-range Bluetooth wireless connection. The solution uses a distributed trust architecture which has the following features:

1. develop and clearly state a security policy;
2. assign credentials to the users and devices (entities);
3. delegate trust to a trusted third party (TTP);
4. give the right to each entity to reason about the user's access rights, which are dynamic;
5. check the credentials of an initiator which should fulfil the policies in order to provide an access control.

The idea is that some authorized person has the right to delegate the use of services in a smart space to another user for a period of time during which the authorized person is in the space. The other user can also give the delegation to yet another user so a chain of delegation will be established. When any user fails to meet the demands associated with a delegated right, the chain will be broken. As a result, no user will be able to do any other action associated with the right. Moreover, the authorized person can perform delegation and revocation in this architecture. In addition, the researchers used trusted XML signatures instead of using X.509 certificates to protect the user's privacy who do not want to log into the systems using their names. A user, who wants to access a service, should first submit their credentials to the security agent which will be responsible for generating the authorization certificate. The user can then use the certificate as a ticket to access specified services or delegate the access to other users. Kagal et al. [24] attempted to solve the problem of access control by using trust, rights and delegation; therefore they developed a flexible scheme of trust used for modelling the permissions and delegations. They considered permissions as the right of an agent and connected rights with actions and according to this the corresponding agent can do a specific action. These permissions can then be extended by delegation from an authorized agent. The authors stated that there is a difference between centralized and distributed systems in authorization. They explained that although there are many different schemes of decentralization (e.g. Access control list ACL and Role-based access control RBAC) which are useful, these are insufficient in providing a design for trust management. According to them, secure systems in general should not only authenticate users, they should also permit users to delegate their rights to other users securely and have a flexible mechanism for controlling this delegation. They claimed that the majority of delegation schemes partially address the issues related to authentication and delegation combined, while others only support authentication, ignoring delegation altogether. There are some schemes that support delegation to some level without providing the flexibility needed, while others do not provide sufficient restrictions on delegation of rights. If a user has sufficient access rights to use a particular resource, then they should have the power to delegate some or all of these rights to others. This should be defined in a security policy, which restricts which rights may be delegated by which agents and to whom. Privileges can be given to trusted agents, who are responsible for the actions of the agents to whom they subsequently delegate the privileges. So, agents will only delegate to agents that they trust. In the work presented in [26], while trust relationships are usually based on identity, they aimed to define a trust model that is capable of modelling scenarios where identity may not be available, yet the context of the scenario is more correctly used in establishing trust relationships. The designed model provides a formal basis for making trust decisions. Examples of such systems include most traditional authentication systems, where trust is established using shared secrets, public/private methods and certificates. They argued the needs for a unified model of trust between entities (people and devices), which has the ability to capture the requirements of the traditional computing world and pervasive computing world. In the traditional computing

205

world, trust is established based on identity, whereas in pervasive computing it is based on identity, physical context or both. Therefore, the authors presented a novel attribute vector calculus which has the ability of modelling trust relationships between entities. The main reason for using such a vector is because it can capture both the context-based and identity-based trust relationships in a simple approach.

The authors in [27] proposed a new trust evaluation model based on the user's past and present behaviour and linked it to a light weight authentication key agreement protocol. They aimed to establish an intermediate phase that evaluates the trustworthiness of connected entities before the service provision phase allows access to resources in smart environments. Their proposed scheme will preserve the user's privacy because it will use non-sensitive information in the evaluation process. They claimed that trust evaluation models are required in smart environments to provide secure yet more flexible environments. Figure 2 shows the trust model architecture, in which judgment is computed based on the reports of experience messages. The indirect trust is achieved by multiplicative relation between judgment and recommendation values given by recommendation messages. In addition, the direct trust is determined through a risk assessment based on a number of positive and negative actions. The final part is the net trust which is a linear combination of both direct and indirect trusts. When the trust values expired, the trust updates occur.

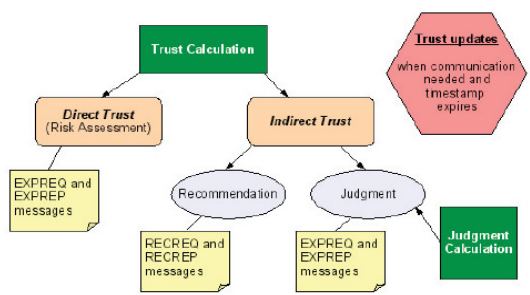

Figure 2 the proposed trust model architecture [27]

From the above review, it can be concluded that there are different ways to establish trust in pervasive computing systems for secure communication among entities. Some authors used context information to allow users to specify their policies depending on the level of mutual trust between two entities. Others proposed alternative approaches from the human and social sciences to distribute trust, where a chain of trust will offer more flexibility in designing policies in pervasive environments. Furthermore, other authors used delegation schemes to provide an access control from an authorized user to another user for a specific period of time. It is evident from the above discussion that pervasive environments require more flexible and dynamic models for defining trust relationships. They also require distributed architectures that are capable of forming and maintaining such relationships.

## 5. Identity in Pervasive Computing

Modern computer systems provide services to multiple users and require the ability to securely identify (authenticate) the users requesting the services. Identifying users is an important step in providing any service, yet it raises a number of issues. For instance, in the case of using Plain Password-based authentication, the password sent across the networks can be intercepted and later used by eavesdroppers to pose as the user, making it unsuitable for use on computer

International Journal of Distributed and Parallel Systems (IJDPS) Vol.3, No.3, May 2012networks, hence the use of encrypted passwords. In addition to the eavesdropping concern, password based authentication is not always convenient; users do not want to enter passwords each time they access a service on the network. This has led to the use of a weaker authentication on computer networks. To overcome these problems a number of systems use stronger authentication methods based on cryptography. When using authentication based cryptography, an attacker listening on the network gains no information that would enable them to falsely claim another person's identity. As authentication provides verification of identity and ensures that the identity declared is really the true identity. This is crucial in enabling access to the right parties. However, it does not describe the privileges entry processes. For instance, a user is authenticated before accessing a database system, but this does not tell the database system which data the user is allowed to access. As we move into the world of pervasive computing, there is an increased transparent interaction between people and smart devices which have computing power. Each entity (user or device) in such environments is continually interacting with hundreds of nearby wireless interconnected computers. As a result, this means that the common security approaches are insufficient to guarantee proper authentication with these entities. Bussard and Roudier [5] pointed out that it is more crucial to "authenticate artefacts" in order to protect the artefact or entity from any possible attack and ensure that these entities are not faked. They concentrated on the authentication of an artefact, which is based on a dedicated challenge-response protocol and merged it with standard security mechanisms to prove that an artefact has some privileges; this is called "local proof of knowledge protocol". They also stated that the verification of an artefact, which knows a secret, can be done by ensuring that it cannot communicate with other devices during the challenge-response. The proposed solution is based on dedicated hardware to ensure quick two-bit message exchange and to circumvent an attack; thus eliminating the need for cryptography algorithms. Moreover they presented a possible attack such as Man-In-the-Middle (MiM) attack to test this solution. They explained how a user can verify that an artefact knows a specified secret based on simple Local Proof Boolean challenge-response protocol and a trusted third party that can provide an evidence for that artefact that it has the required rights. Furthermore a time-based solution (based on Round Trip Time measurement) was proposed based on hardware architecture, where exchanging messages between the user and the artefacts takes place at the physical layer.

Al-Muhtadi et al. [28] used context and location information based on encryption to achieve their main goal which is finding an efficient authorization mechanism. They developed a distributed middleware called 'Gaia', which can be run on top of existing operating systems in order to provide a basic structure (substructure) and central services for fabricating a general purpose pervasive computing environment. The researchers developed a framework that provides efficient key management for 'secure group communication' based on defining context. The framework is based on two-layer encryption, one for group membership and the other for location information. The two layers use the Advanced Encryption Standard (AES) 128-bit and 256-bit crypto algorithm. The purpose of these layers is to forbid an unauthorized device to eavesdrop or spy on an event when it is not in the target region. The researchers defined the target region as "spatial region and corresponding files created by an administrator or an authorized entity, so that only users located within that region can read these files". Using these layers, a device, that is authorized but not located within the target region, will be unable to penetrate the outer encryption layer. In such a scheme, the second layer (location encryption layer) is stripped only after an entity's location has been confirmed. Therefore, a group member who leaves the active space is unable to decrypt the data with its group key because of the existence of the location encryption layer. Hence, re-keying of the remaining group members is not essential. When an entity's permissions are changed, only the keys for the affected levels are revoked. In such a case the re-keying includes producing a new group key and then distributing this key to that level or group. In [29] authors used Gaia to provide the necessary core services to support and manage active spaces and the pervasive applications that run within these spaces. By using Gaia, it is

207

possible to construct an active space where authors incorporate a number of authentication mechanisms; each mechanism attains a specific value known as the 'confidence value'. This value ranges from 0 to 1 depending on the device and protocol used in the authentication process. In order to increase the confidence value, a specific authentication mechanism may include any number of authentication processes. Reasoning techniques are used to formulate the net confidence value from the partial confidence values. This authentication provides a unique feature which decouples the authentication procedures and authentication devices into two sections. The first is an Authentication Mechanism Module (AMM) encompasses all the authentication procedures available such as challenge-response, Kerberos, SESAME, etc.. The second is an Authentication Device Module (ADM), which is device-dependent, is used for each authentication device such as PDA, smart badge, etc. This decoupling facilitates the incorporation of a new protocol in the AMM section or a new module in the ADM section for a new authentication device without interacting with the other section.

Lee et al. [30] described a process for distributing the key which requires each individual entity in a group to re-authenticate itself before receiving a new key. The process of re-authenticating checks and ensures the authenticity of the authorization method and hence adds an overhead to the management entity. This is compared to the multi-layer encryption method that was presented in [29], which eliminates the requirement for re-authenticating entities within a group when a contextual change causes changes to the permissions of the group.

Seigneur et al. [31] claimed that entity recognition is more universal than authentication schemes and pervasive computing environments can make use of this approach. They based their claim on the fact that devices in pervasive computing environments have the ability to connect to systems on a point to point basis, control each other and become aware when connecting to each other and exchange information among them. They claimed that the recognition of previous interaction to choose whether to co-operate or not, is more important than the Identity. They believed that using an entity recognition approach is better than using the traditional authentication approaches such as Kerberos, PKI, etc, because entity recognition strategies do not always need a human interface phase, but may need a process to raise awareness of the surrounding entities depending on their estimated importance. Therefore, they proposed an approach called "pluggable recognition module (PRM)" which is compatible with authentication approaches and is pluggable to allow various applications to choose different schemes to identify collaborator (partners). Moreover, they showed that instead of knowing "whom does this entity represent", we should know "can we recognize that entity as a trusted collaborator regardless of whichever it represents". Thus they introduced virtual pseudonymity (which means no need for distinguished names) and therefore they did not require the ability to establish the real identity of a given entity. Instead, they simply required the ability to recognise other entities, e.g. through their name, location, digital signatures or other means. Furthermore, they tried to limit the human interface in the smart environment by using their approach instead of using the authentication schemes. They also identified many requirements for their design. For example, they identified that to recognize an entity there must be no demand for on-line interaction with a central host. This scheme may also rely on third-parties to achieve higher scalability.

Stajano[32][33] presented a security policy model called "Resurrecting Duckling model" which is an example of entity recognition, in which the main idea is that a Slave device can pose to be a Master by the transfer of an imprinting key. This model describes the properties that a system should possess to implement a satisfactory secure transient association using imprinting to establish a shared secret. They applied their duckling model to a typical scenario to identify and pair a remote control unit with a single user, yet retain a certain amount of flexibility. When a user has a universal remote control unit to control different devices at home, the user needs to ensure that the device will obey their commands, not those of a neighbour's. Also, when a device breaks down, the user needs to retain control of the other devices. The model is based on the scenario of a duckling (device) which identifies and fallows its mother (a user) as the first entity which sends a learning process key or imprinting key (a duckling identifies the first moving entity it sees and hears as its mother). However, what would happen when the duckling (device) dies or breaks down? According to the Resurrecting Duckling model, a duckling that dies can become alive again (resurrected) later with a different soul. The authors treated the device as the body and the shared secret (software) like a soul. Only the owner has the authority to end the Duckling's life, which will return to its pre-birth status and can accept a new imprinting key. Moreover, if the damage was un-repairable, the manufacturer (who has the master password) has the right to command the device to die. To achieve this goal the researchers showed that the devices should be designed based on tamper resistance that will help to destroy the duckling without affecting its body. In addition, they extended their idea by allowing the duckling to connect to other peers and other mothers in addition to their original mother and to give a duckling the ability to send orders to other ducklings. This, however, will lead to more complex situations. They introduced a two-level policy to solve the problem, the mother policy and the duckling policy. The mother has the control and edits the duckling (device) policy and also can delegate her control to another mother, but the duckling decides what certification must be shown to perform a specific task. The proposed security policy model (Resurrecting Duckling policy model) can be used to solve the problem of "secure transient association" and can reduce the risks of such a system.

Creese et al. [34] indicated that the traditional concepts of entity authentication are inappropriate to pervasive computing environments and presented their view on how to modify and change these notions to solve many problems in such an environment. They argued that the contextual attributes such as location and manufacturer's certificates should be valid to establish an accurate level of assurance. Their work focused on establishing trust as opposed to handling dynamically changing permissions. They stated that unlike traditional authentication which is based on password, token, biometric and public key encryption, in pervasive computing environments there is no need for pre-existing list of trusted parities. The trustworthiness is based on the device attributes such as location and human contact not on identity. Moreover, in order to create a good basis for trust (assurance) one should have sufficient evidence of legitimacy of the devices (such as knowing the vendor and the manufacturer). They argued the need for a flexible and dynamic security policy for the pervasive environment and invented a metric graph for comparing authentication sets of attributes and to help making decisions based on a suitable policy. Zia et al. [3] gave a special significance to the need of risk management which compromises threat analysis. They qualified their claim on the basis of the wide flexibility in the interconnection between many different devices and how this may be achieved through context-based access control mechanisms. In traditional computer access control, access privileges are based on the user's identity and a pre-defined list of access controls (ACL) outlining a certain authority to the user. In contrast, in pervasive computing environments authorisation is not based on identity only, which then reflects the user's requirements to use a specific service. This is a result of the wide spread use of context-based devices. Their objective was to formulate a practical risk management technique in environments, where services are accessible to and consumed by processes in order to perform business needs. This technique is different from a risk prevention methodology which causes limited interactions and will in turn decrease business effectiveness. Thus various risk models may be adopted by individual risk owners according to their risk appetite. These result in differing levels of security and communication and therefore business process efficiency. As outlined before, authentication is crucial to pervasive systems because devices may communicate with untrusted or unknown entities. Designing a standard authentication process in a pervasive environment where devices have mobility and transparent interaction with users would be difficult to implement. Therefore Li et al. [35], proposed an authentication protocol for secure use of public information within such an environment but without the need to access a trusted third party (TTP). Their protocol can prevent the passive (just listening) and active (with control to modify and drop data) attacks, by establishing a new PKI and new signature scheme. The proposed protocol can be implemented using symmetric or asymmetric cryptography algorithms. They identified the main properties which should be provided in an authentication protocol that makes use of the public utilities as:

- Entity authentication which means the user should check which utility they are interacting with;
- Data confidentiality which means the user should be certain that the transmitted information should be encrypted so that no attacker (passive or active) can reveal the information.

Garzonis et al. [36] stated that in user interaction the most important issue of security is the identification and authentication which means asking the user to enter his/her user name and password. However, in pervasive environments, identification and authentication using a user-remembered password raises the problem of usability and vulnerability to attacks. Moreover, in traditional network security the process of identification and authentication can be done using some identification information such as the IP address of the user's device. However, in pervasive environments the user is expected to use many devices in different networks, requiring a new method to identify the user instead of relying on the device being used. The researchers proposed a mixture of embedded biometrics and the use of IPv6 header extensions for the interaction between network and human, rather than computing devices. This scheme can provide a personalized interaction and secure identification procedure for the user according to their preferences. By using biometrics systems the re-authentication process can be done without the need for the user's interference and without the need for entering a password. Therefore, re-authentication using biometric systems can solve this problem, by making the user information as part of each data packet leaving the device. For that reason, they used IPv6, which includes a header extension known as an "option mechanism". IPv6 options are placed in separate extension headers that are located between the IPv6 header and the transport-layer header in a packet [37]. Most IPv6 extension headers are not examined or processed by any router along a packet's delivery path until the packet arrives at its final destination. For this reason and others, IPv6 options may be used to include authentication and security encapsulation options; such options were not possible in IPv4. A good example of this is the insertion of biometric systems as authentication of information to link information to users. They also showed that this mechanism can provide a level of support for context awareness by carrying information about the device being used and the characteristics of the user. This information will remove the need for transmitting contextual information separately, and changes to the context will be updated dynamically through this network protocol. The main idea of the combination of their research is to decrease the inverse relationship between security and usability and allow the user to perform secure activities in this environment without effecting users' privacy.

The research in [38] argued that due to the characteristics of pervasive computing environments there is a challenge in asserting the user's identity. In such environments, it is impractical for users to prove their identities (through authentication) every time they cross or move among various networks. So, to reduce the burden of frequent user authentication and verify user identity with a level of certainty, an approach is required for verifying the user identity when interacting within a smart environment. The authors suggested developing a new non-intrusive and adaptable technique for asserting the user identity. The proposed system called Non-Intrusive Assertion System (NIAS) becomes aware of the user's intention and behaviour while attempting to verify their identity and maintaining confidence on the identity of the user. NIAS has the ability to monitoring certain aspects of the user's behaviour.. The system then uses the user behaviour, assert the user's identity.

Most authors argued that conventional security techniques are insufficient to securely establish the identity of users in pervasive computing environments. Some research has presented modified techniques such as the use of contextual attributes, recommendation from a trusted third party (TTP), or a combination of IPv6 and embedded biometric systems. In our opinion, establishing the identity of a user that interacts with the environment remains a crucial part of the overall security provisioning in pervasive environments. As these environments dynamically change their behaviour according to the current situation of the user (which is known as context), it is essential that the environment would identify its users securely and reliably. Furthermore, using a password-based security mechanism (or any other mechanism that depends on direct response from the user) to identify the user is proven to cause a usability problem, and may reduce the overall flexibility and workability of the environment. However, users, as well as devices, need to be assured that identity is securely established and not mistaken with someone else's before any meaningful communication can take place between the user and the environment. We believe that identification of users in such environments will remain as an open research issue over the next few years, which will attract a lot of attention among the research community in this area. We also believe that this issue poses further challenges to the adoption of such environments, similar to those that are posed by privacy. For example, in a medical environment, it is paramount that users are given enough assurance that the identity of patients is securely maintained at all times and not mistaken with other patients' identity.

## 6. Comparison between Traditional Computing and Pervasive Computing

As discussed in the previous sections, there are numerous differences when considering security related issues in the design of pervasive and traditional computing systems. We believe that it is important to highlight the differences between the two types of system, as this will make it easier for researchers to appreciate the corresponding requirements. Tables 1 to 5 summarise these differences with respect to a number of criteria; in each criterion a number of differences may be found. The tables use the following categories for the purpose of comparison:

1. User Interaction: This criterion compares issues related to the relationship and interaction between the user and the computing devices, how connection is established between networked devices and risk management when associating users to devices. See table 1
2. Access control: This criterion compares issues related to controlling access to resources, authorisation procedures, policies and security architectures. See table 2
3. Trust: This criterion compares issues related to how trust relationships are established, the role and nature of trust relationships and the complexity of such relationships. See table 3.
4. Privacy: This criterion compares issues related to users' perception and confidence in the way their information is stored and exchanged. It also compares issues related to the degree of risk in compromising the privacy of user information. See table 4.
5. Identity: This criterion compares issues related to the way the user identity is established and the risks involved when such identity is falsified or misused.

Table 1. Comparison between Traditional and Pervasive Computing Environments with respect to User Interaction

| Criterion | Traditional Computer Networks | Pervasive Computing Environments |
|---|---|---|
| User Interaction | ▪ In traditional systems, the user's intervention is considered necessary for initiating the connection to a specific device within a network. | ▪ In pervasive systems devices will be connected automatically to a network without the users' participation. The devices will initiate the connection for them and the users don't need to know which network they have accessed [39]. |
|  | ▪ Traditional network systems have a well defined process for risk assessment and procedures. | ▪ In pervasive systems more flexible and varying risk assessment process is needed, due to the unpredictable and highly distributed interaction [1]. |
|  | ▪ In traditional systems, standard authentication protocols such as Kerberos, IPSec, and SSL, are used for controlling access to network resources. | ▪ Standard authentication protocols cannot be readily used pervasive system environments, because they can provide the required mobility and scalability needed such environments [24]. |
|  | ▪ The computing model is based on localised desktop devices, where there is one–to–one relationship between the machine and user. | ▪ Computing is highly distributed into the surroundings and onto the user's personal digital devices. There is a many–to-one relationship between the machine and user [1][40]. |
|  | ▪ The desktop devices are fully controlled by the user. | ▪ The user has less control on the device actions. |
|  | ▪ The user is conscious about the interaction with the desktop devices, where all information supplied by the user is under full control of the user at all times. | ▪ The user is unconscious about the interaction with many devices (the number of devices may not be known) and the connection between these devices will be unknown. The user is unaware of what information is sent to what device at any particular point in time [9]. |

Table 2. Comparison between Traditional and Pervasive Computing Environments with respect to Access Control

| Criterion | Traditional Computer Networks | Pervasive Computing Environments |
|---|---|---|
| Access control | ▪ Access to resources can be done by knowing the identity of a device such as its IP address or MAC address. | ▪ The user needs the ability to access any resource and service at any time from any place without a need to know the identity of the device [1]. |
|  | ▪ Traditional authorisation or access privileges have been based on the user's identity, where a pre-defined list of identities is maintained for authorised access to a resource or service. Using the identity, the system issues a ticket detailing authorisation for accessing resources on the network. | • User identity alone is insufficient to control access to resources and services. A more sophisticated access control mechanism is needed to grant access based on user identity, context and behaviour. This is due to the nature of the environment where users' needs are more dynamic and the services are constantly changing [25]. |
|  | ▪ Security architecture in traditional computing systems involves using firewalls to restrict the access to network resources. | ▪ Using firewalls in pervasive computing will not be effective because the network architecture is more complex, and may comprise multiple domains. This may require installing firewall software within each device [41]. |
|  | ▪ Traditional security architectures use centralized authorization servers to grant a | ▪ In pervasive computing systems, more distributed security architectures are needed, |
| | user access to a resource within the network. | such as keeping uniform security policies across the distributed components of the architecture [23]. |
| | ▪ Security policies are usually static which are typically based on layer 3 or 4 information. | ▪ Dynamic policies, that take into consideration the privacy of the user's sensitive information, are required because of the flexibility of using many different devices which might work in different networks. Moreover, both the devices and applications will be constrained by the limits of available resources such as communication capabilities, computation and storage [42]. |

Table 3. Comparison between Traditional and Pervasive Computing Environments with respect to Trust

| Criterion | Traditional Computer Networks | Pervasive Computing Environments |
|---|---|---|
| Trust | ▪ In traditional networks, trust relationships are established based on identity, recommendation from third trusted party (TTP) or reputation. | ▪ In pervasive computing, trust relationships are established using the identity of a user and their context information (behaviour and attributes) [26]. |
| | ▪ Trust in traditional systems is a means for controlling access to resources. | ▪ In pervasive environments, trust is more a general term as it also includes a measure of how accurate the information is. |
| | ▪ Trust relationships in traditional systems are static in nature; once the relationship is formed between the trustor and trustee, it remains valid until it is broken explicitly by the trustor. | ▪ In pervasive systems, the relationship is more dynamic and based on historical information and risk assessment. Every time a device requires access to a resource, the trust relationship is re-assessed according to the device's current and previous status. |
| | ▪ In traditional systems, trust relationships are simple, and include two parties: trustor and trustee. | ▪ In pervasive systems, more complex relationships may be formed by delegating trust from one user to another to form a chain of trust relationships [26]. |

Table 4. Comparison between Traditional and Pervasive Computing Environments with respect to Privacy

| Criterion | Traditional Computer Networks | Pervasive Computing Environments |
|---|---|---|
| Privacy | ▪ In traditional networks, privacy is less problematic as despite people are concerned about holding and storing their personal information, they know where this information is stored and used. | ▪ In pervasive computing environments, privacy is more significant, for people are less willing to exchange their personal information with the environment. This is because they are unaware or unsure where their information being held and used [1]. |
| | ▪ There is a lower risk in storing personal information on traditional networks, as they can only be accessed by authorized users. | ▪ There is a higher risk involved in storing personal information on pervasive and mobile systems, as they may be accessed anywhere and by anyone [1]. |

International Journal of Distributed and Parallel Systems (IJDPS) Vol.3, No.3, May 2012Table 5. Comparison between Traditional and Pervasive Computing Environments with respect to Identity

| Criterion | Traditional Computer Networks | Pervasive Computing Environments |
|---|---|---|
| Identity | ▪ In traditional systems, the user identity is established and verified by using the common authentication methods, such as checking a password, swiping a smartcard, or other means of proving that the user is who they claim to be. | ▪ In pervasive computing environments, more subtle ways are required to establish the user identity, because common authentication protocols may not be adequate [24]. |
| | ▪ The risk of identity theft is mainly linked to stealing a password or credentials which an attacker will use to impersonate someone else. | ▪ The risk of identity theft is higher in pervasive computing because there is a higher risk of losing the user's device; such as PDA or a mobile phone, where identity is normally stored. |

## 8. Future Research Directions

In this section, we explore the directions and topics of research related to privacy, trust and user identity in pervasive computing environments.

### 8.1 Research Directions Related to Privacy in Pervasive Computing Environments

Privacy is an important issue within pervasive computing, as people become concerned when they are unaware where their personal information is being saved, by whom and for what purpose it would be used. To address these concerns, possible research areas in this direction are highlighted below. These areas will provide users with greater confidence about the way in which their information is being utilised when interacting with pervasive computing environments.

- Investigating and developing new privacy architectures which can measure how much of the user's personal information should be given and determine which part of this information needs to be collected by the environment for a particular purpose. These architectures will provide users in pervasive environments with greater control in the way in which their information is being exchanged.
- Extensions of the anonymity and pseudonymity concepts to prevent the leakage of personal information in pervasive environments. Such extensions will protect the users' privacy and provide them with flexibility by giving the rights to users to choose whether to distribute and exchange their personal data or not. One possible extension is to provide multiple levels of anonymity, in which the users could make informed choice as a trade-off between privacy and functionality.
- Designing abstraction models and policy languages to control privacy management. Such models can be implemented by extending Langheinrich's principles [9], which are: Notice, Choice and Consent, Proximity and Locality, Anonymity and Pseudonymity, Security and Access and Recourse. New models and languages are needed to enable designers to easily define policies for managing user privacy based on these policies.
- Developing privacy policies for the exchange of personal information based on interpreting and abstracting users' contextual information. Contextual information may be used to provide insight to the system for deciding which parts of the user information is needed (for a specific functionality) and hence retrieve the relevant information without having to exchange the irrelevant parts. Research would be needed to define dynamic policies for this purpose.

### 8.2 Research Directions Related to Trust in Pervasive Computing Environments

Because of the dynamic nature of pervasive computing environments, forming trust relationships presents a challenge to designers. To overcome such a challenge possible research topics include:

- Designing distributed flexible and dynamic trust architectures which have security policies that could focus on providing more control over the access of services. These architectures will be capable of supporting, forming and maintaining trust relationships dynamically.
- Designing adaptable protocols which should maintain a high scale of user mobility in pervasive environments. These protocols should have the ability to exploit localized trust establishment and decision making.

### 8.3 Research Directions Related to Identity in Pervasive Computing

Establishing user identity within pervasive computing environments requires innovative methods for authenticating users which take into consideration the dynamic and mobile nature of such environments. To deal with this issue, possible research topics include:

- Implementing new techniques to provide non-intrusive mechanisms to authenticate users as they interact with a huge number of entities in pervasive computing environments. Such approaches would shift users away from the classical intrusive techniques in traditional network systems (for example, entering a password) towards non-intrusive techniques. In addition, such approaches will help achieve a balance between security and usability.
- Designing new architectures and protocols which can cope with changing identity across different domains. As users in pervasive environments have many interactions with various devices and applications, research is needed for designing new architectures and protocols to manage multiple identities of users as they cross domain boundaries.
- Developing new adaptable protocols and models for sensing, gathering and filtering user contextual information. Acquiring contextual information such as user's activity or behaviour will help in verifying the correct identity. These protocols need to support multiple personas (characters) of a single user in pervasive computing domains and maintain the objectives of pervasive computing in creating seamless environments and delivering distributed services. In meeting such adaptable protocols and models, this will support the mobility and dynamism features in pervasive computing environments.

## 9. Summary

This paper has reviewed a number of technical challenges related to designing secure pervasive systems and compared them to more traditional computing environments. A number of research papers have been reviewed to cover various challenges and technological advances in the subject. The major differences between traditional computer networks and pervasive computing environments have been highlighted. In pervasive environments, issues related to assessing resources are similar to peer-to-peer communication issues. Users in such environments will have the ability to gain access to any resource/service at anytime from anywhere. This fact will result in serious implications since devices are constantly interacting with other devices outside their (home) environments. This new smart world will bring many differences in comparison with recent traditional computer systems. Generally, policing resources in traditional computing systems means using firewalls for access control, static policies, and tendency to focus on network layers and static risk assessments. While in pervasive computing systems, access control is based on using authentication, identity management and trust because such systems are more distributed, dynamic and have a high risk as users' personal information would be accessed from anywhere by anyone.

We believe that authentication in a pervasive computing environment should be considered as the first stepping stone, because it is important to reliably establish the identity of the user in such environments. We also believe that authentication requires the development of new techniques and policy systems that also take into account the user's contextual information.

Privacy, trust, and identity are highlighted as the main considerations in the design of pervasive environments in comparison to more traditional computing systems. It is vital to provide solutions for these issues in pervasive computing in order to be truly beneficial and socially acceptable and for the users to take part comfortably within such trustworthy environment. Privacy is a major challenge in pervasive computing when compared to traditional computing systems. For that reason, a number of proposed solutions have used concepts such as anonymity and pseudonymity which can also imply trust, to prevent the leakage of personal information. Trust allows a greater flexibility in designing security policies and providing more control over the accessed services. Trust relationships are mainly established using context information such as the behaviour and attributes of a user. When a user establishes more than one connection with different devices, each connection needs a verification process. The aim of the verification process is to confirm precisely the identity of the person and enable the pervasive system to cope with a stolen identity. Different approaches have been surveyed in the identity section. It is apparent that acquiring contextual information about a user and saving them as history could be used to authenticate the user and will help in verifying the correct identity. However, the verification process is still prone to security attacks as it is exposed to more devices and external networks.

Many research papers in pervasive environments pay attention to the risk involved in user identification and authentication; this makes identity and trust in such environments very closely related. Establishing user identity requires a measure of trust (accuracy) in the process of verification as an indication of the system confidence in the user established identity.